\documentclass[prl,twocolumn,preprintnumbers]{revtex4-1}%
\usepackage{epsf,epsfig}
\usepackage{amssymb,amsmath,amsfonts}
\usepackage{color}
\usepackage{braket}
\usepackage{subcaption}
\def\pb[#1,#2]{\{#1, #2\}}
\def\deb[#1,#2]{[#1,#2]_{\text{D.B.}}}

\def\tO{\widetilde{\cal O}}

\def\tphi{\widetilde{\phi}}

\def\Or[#1]{{\text{O}}\left({#1}\right)}
\def\dotl[#1,#2]{\left\langle #1,\, #2 \right\rangle}
\def\dotlb[#1,#2]{\left\langle #1,\, #2 \right\rangle}
\def\dotlm[#1,#2]{\left[ #1,\, #2 \right]}
\def\dotp[#1,#2]{(\vect{#1} \cdot\vect{#2})}
\def\aff[#1,#2]{\hat{#1}(#2)}

\def\n4sym{{\cal N}=4 SYM}
\def\>{\rangle}
\def\<{\langle}
\def\weight[#1,#2,#3]{\{(#1),#2,#3\}}
\def\ads[#1]{$\text{AdS}_{#1}$}

\def\tarelr[#1]{\widetilde{a}^{\text{rel}}_{R#1}}
\def\Oright[#1]{{\cal O}_{R#1}}
\def\Oleft[#1]{{\cal O}_{L#1}}
\def\aleft[#1]{a_{L#1}}
\def\arelr[#1]{a^{\text{rel}}_{R#1}}

\newcommand{\be}{\begin{equation}}
\newcommand{\ee}{\end{equation}}
\newcommand{\beq}{\begin{equation}}
\newcommand{\eeq}{\end{equation}}
\newcommand{\ben}{\begin{displaymath}}
\newcommand{\een}{\end{displaymath}}
\newcommand{\beqa}{\begin{eqnarray}}
\newcommand{\eeqa}{\end{eqnarray}}
\newcommand{\bea}{\begin{eqnarray}}
\newcommand{\eea}{\end{eqnarray}}
\newcommand{\bean}{\begin{eqnarray*}}
\newcommand{\eean}{\end{eqnarray*}}
\newcommand{\ba}{\begin{array}}
\newcommand{\ea}{\end{array}}
\newcommand{\bi}{\begin{itemize}}
\newcommand{\ei}{\end{itemize}}

\newcommand{\tfd}{|\Psi_{\rm \:\!\!  TFD}\rangle}
\newcommand{\tfdbra}{\langle\Psi_{\rm \:\!\! TFD}|}
\newcommand{\bs}{\begin{split}}
\def\sess\end{split}

\newcommand{\vect}[1]{{#1}}

\def\tO{\widetilde{\cal O}}

\def\rsz{|\Psi_{0}\rangle}
\def\lsz{\langle \Psi_0|}

\def\tO{\widetilde{\cal O}}

{}{}
{}{}
\def\vereq#1#2{\lower3pt\vbox{\baselineskip1.5pt \lineskip1.5pt
\ialign{$\m@th#1\hfill##\hfil$\crcr#2\crcr\sim\crcr}}}
\makeatother

\begin{document}

\title{On the interior geometry of a typical black hole microstate}

\preprint{CERN-TH-2018-094}

\author{Jan de Boer}
\email{j.deboer@uva.nl}
\affiliation{Institute for Theoretical Physics, University of Amsterdam, Science Park 904,
1098 XH Amsterdam, The Netherlands}

\author{Rik van Breukelen}
\email{rik.van.breukelen@cern.ch}
\affiliation{Theoretical Physics Department, CERN, CH-1211 Geneva 23, Switzerland}
\affiliation{Geneva University, 24 quai Ernest-Ansermet, CH-1214 Geneva 4, Switzerland}
 \author{Sagar F. Lokhande}
\email{sagar.f.lokhande@gmail.com}
\affiliation{Institute for Theoretical Physics, University of Amsterdam, Science Park 904,
1098 XH Amsterdam, The Netherlands}

\author{Kyriakos Papadodimas}
\email{ kyriakos.papadodimas@cern.ch \vspace{0.4cm}} 
\affiliation{Theoretical Physics Department, CERN, CH-1211 Geneva 23, Switzerland}
\affiliation{Van Swinderen Institute for Particle Physics and Gravity, University of Groningen, Nijenborgh
4, 9747 AG Groningen, The Netherlands}
 \author{Erik Verlinde}
\email{e.p.verlinde@uva.nl}
\affiliation{Institute for Theoretical Physics, University of Amsterdam, Science Park 904,
1098 XH Amsterdam, The Netherlands}

\begin{abstract}
\noindent
We argue that the region behind the horizon of a one-sided black hole can be probed by an analogue of the double-trace deformation protocol of Gao-Jafferis-Wall. This is achieved via a deformation of the CFT Hamiltonian by a term of the form 
${\cal O} \widetilde{\cal O}$, where $\widetilde{\cal O}$ denote the state-dependent ``mirror operators''. We argue that this deformation creates negative energy shockwaves in the bulk, which allow particles inside the horizon to escape and to get directly detected in the CFT. This provides evidence for the smoothness of the horizon of black holes dual to typical states. 
We argue that the mirror operators allow us to perform an analogue of the Hayden-Preskill decoding protocol. Our claims rely on a technical conjecture about the chaotic behavior of out-of-time-order correlators on typical pure states at scrambling time. 
\end{abstract}
\maketitle

\section{I. Introduction}

The black hole information paradox is 
related to the question of smoothness of the black hole horizon \cite{Mathur:2009hf,Almheiri:2012rt}. The latter question becomes particularly sharp for typical CFT states dual to a large black hole in AdS. It is challenging to reconcile the smoothness of their horizon to unitarity of the dual CFT \cite{Almheiri:2013hfa, Marolf:2013dba}, even though these black holes do not evaporate. 
In \cite{Papadodimas:2012aq,Papadodimas:2013b,Papadodimas:2013} it was argued that these problems can be resolved by describing the space-time behind the horizon  using state-dependent CFT operators, which are partly selected by their entanglement with fields in the exterior. A related proposal from a somewhat different perspective was described in \cite{Verlinde:2012cy, Verlinde:2013uja}. It remains a challenge to fully understand the geometry dual to a typical black hole microstate.

In \cite{Gao:2016bin}, it was realized how to probe the horizon of a two-sided eternal AdS black hole by using double-trace deformations of the CFT Hamiltonian. 
This protocol, reviewed in the next section, has provided evidence for the smoothness of the eternal black hole and  the ER=EPR proposal \cite{Maldacena:2013xja}. It was further discussed in \cite{Maldacena:2017axo, vanBreukelen:2017dul, Bachas:2017rch} and applied to a class of a-typical pure states in \cite{Kourkoulou:2017zaj, Almheiri:2018ijj,Brustein:2018fkr}.

In this paper we develop a similar protocol for one-sided black holes dual to typical pure states in the CFT. This protocol relies on perturbing the Hamiltonian by state-dependent operators and allows us to connect the smoothness of the horizon 
of a typical pure state to properties of CFT correlators. Moreover, it provides an explicit CFT realization of an analogue of the Hayden-Preskill protocol \cite{Hayden:2007cs}.  More details will be provided in upcoming work \cite{long}.

\section{II. Two-sided black hole}

The thermofield double state, which is holographically dual to an eternal two-sided AdS black hole \cite{Maldacena:2001kr}, is an entangled state in the tensor product of two identical CFTs (called ``left'' and ``right''),
\begin{equation}
\tfd= {1 \over \sqrt{Z(\beta)}}\sum_{E} e^{-\frac{\beta E} {2}} \ket{E}_L \otimes \ket{E}_R,
\end{equation}
where $\beta$ is the inverse temperature and we sum over energy eigenstates.
The two CFTs are not interacting, therefore operators on the left and right commute $[\mathcal{O}_L,\mathcal{O}_R]=0$ and no information can be transferred between the  CFTs. Equivalently, in the bulk the Einstein-Rosen wormhole  is not traversable.
 
In \cite{Gao:2016bin} it was argued that the wormhole can become traversable if we couple the two CFTs by a double-trace interaction of the form
$ V= \mathcal{O}_L(0)\mathcal{O}_R(0)$ which then allows for geometric transfer of information,
if the sign of the coupling is appropriately chosen.
An example of a CFT correlator which can diagnose traversability is \cite{Maldacena:2017axo}
\begin{align}
\label{tfdcor}
C \equiv 
 \tfdbra [\phi_L(-t),  e^{-i g V} \phi_R(t) e^{i g V}]\tfd.
\end{align}
A probe is created on the left by $\phi_L(-t)$ and detected on the right by $\phi_R(t)$.
 Without the double trace interaction $V$, we would have $[\phi_L(-t),\phi_R(t)] = 0$. When including $V$ certain terms in \eqref{tfdcor} grow exponentially with $t$,  as  typical  for  out-of-time-order  commutators  in  chaotic  systems \cite{Maldacena:2015waa}. 
Around scrambling time $t = {\beta\over 2\pi} \log S$, we see \cite{Maldacena:2017axo} a signal in the correlator \eqref{tfdcor} representing the probe crossing the wormhole, thus demonstrating smoothness of the horizon of the two-sided eternal black hole.

\section{III. One-sided black hole}

We consider a typical state in a large $N$ holographic CFT, which can be thought of as a random superposition of energy eigenstates
\be
|\Psi_0 \rangle = \sum_{E_i \in (E_0,E_0+\delta E)} c_i \, |E_i\rangle,
\ee
from a narrow energy band of width $\delta E \sim O(N^0)$ and $c_i$ are randomly chosen with the uniform Haar measure. We take $E_0$ to be in the regime dominated by a large AdS black hole in the bulk.

These are almost time-independent equilibrium states. The bulk dual contains at least the exterior of the black hole. It has been proposed \cite{Papadodimas:2012aq,Papadodimas:2013b,Papadodimas:2013} that the interior can be described
using the ``mirror operators'', denoted as $\widetilde{\cal O}$. These operators play a role similar to ${\cal O}_L$ in the two-sided black hole and we will use them to perform an analogue of the experiment discussed in the previous section.

We will now review the mirror operator construction. First we define a ``small algebra'' ${\cal A}$ corresponding to simple observables in effective field theory. Then, 
given a typical black hole microstate $\rsz$, we define the ``small Hilbert space'', also called code-subspace, as
\be
{\cal H}_{\rsz} = {\rm span} \{ {\cal A} \rsz\}.
\ee
This subspace is the one relevant for describing effective field theory in the bulk.

If $\rsz$ is a black hole microstate, it follows \cite{Papadodimas:2012aq,Papadodimas:2013b,Papadodimas:2013}  that 
the representation of the algebra ${\cal A}$ on the subspace ${\cal H}_{\rsz}$ is reducible and the algebra has a non-trivial commutant ${\cal A}'$. The commutant ${\cal A}'$ can be concretely identified by an analogue of the Tomita-Takesaki construction \footnote{See \cite{Haag:1992hx} for a physics-motivated introduction,  \cite{Papadodimas:2012aq,Papadodimas:2013b,Papadodimas:2013} for application to the black hole interior and \cite{Witten:2018zxz} for a recent review of applications in QFT.} and it is natural to associate ${\cal A}'$ with the left  region of the extended AdS-Schwarzschild solution.

Following this, we can define the mirror operators on the code subspace to act as
\begin{align}
 \begin{split} \label{mirrordef}
 \tO_\omega \rsz  &= e^{-{\beta H \over 2}} {\cal O}_\omega^\dagger e^{\beta H\over 2}\rsz,  \\
 \tO_\omega {\cal O}_{\omega_1}...{\cal O}_{\omega_n} \rsz & = {\cal O}_{\omega_1}...{\cal O}_{\omega_n} \tO_\omega \rsz, \\
 [H,\tO_\omega]{\cal O}_{\omega_1}...{\cal O}_{\omega_n} \rsz &= \omega \,  \tO_\omega {\cal O}_{\omega_1}...{\cal O}_{\omega_n} \rsz. 
 \end{split}
\end{align}
Here ${\cal O}_\omega$ denote the Fourier modes in time of single-trace operators.
The extension of the operators on the rest of the Hilbert space is irrelevant for the following calculations. We notice that $[{\cal O},\tO]=0$ only inside the code subspace and may be nonzero as an operator. 
 
 While the definiton of $\tO$ to subleading orders in $1/N$ is not unique, and in particular it will be related to the details of gravitational dressing of local bulk operators, for the purposes of this paper we extend these equations even when we include $1/N$ effects. 

The equations \eqref{mirrordef} are defined only for modes with $|\omega|< \omega_*$ where $\omega_*$ is a large, but $N$-independent frequency. Because of this restriction, it is not meaningful to define the mirror operators for sharply localized operators ${\cal O}(t)$. Moreover, the time argument of {\it smeared} mirror operators in position space is assigned so that $\lsz O(t_1) \tO(t_2)\rsz$ depends on $t_1+t_2$.  This also determines the time-ordering of mirror operators. We emphasize that these operators are explicitly time-dependent, as will be discussed in more detail in \cite{long}.
 
\begin{figure}[!t]
\begin{center}\includegraphics[width=.2\textwidth]{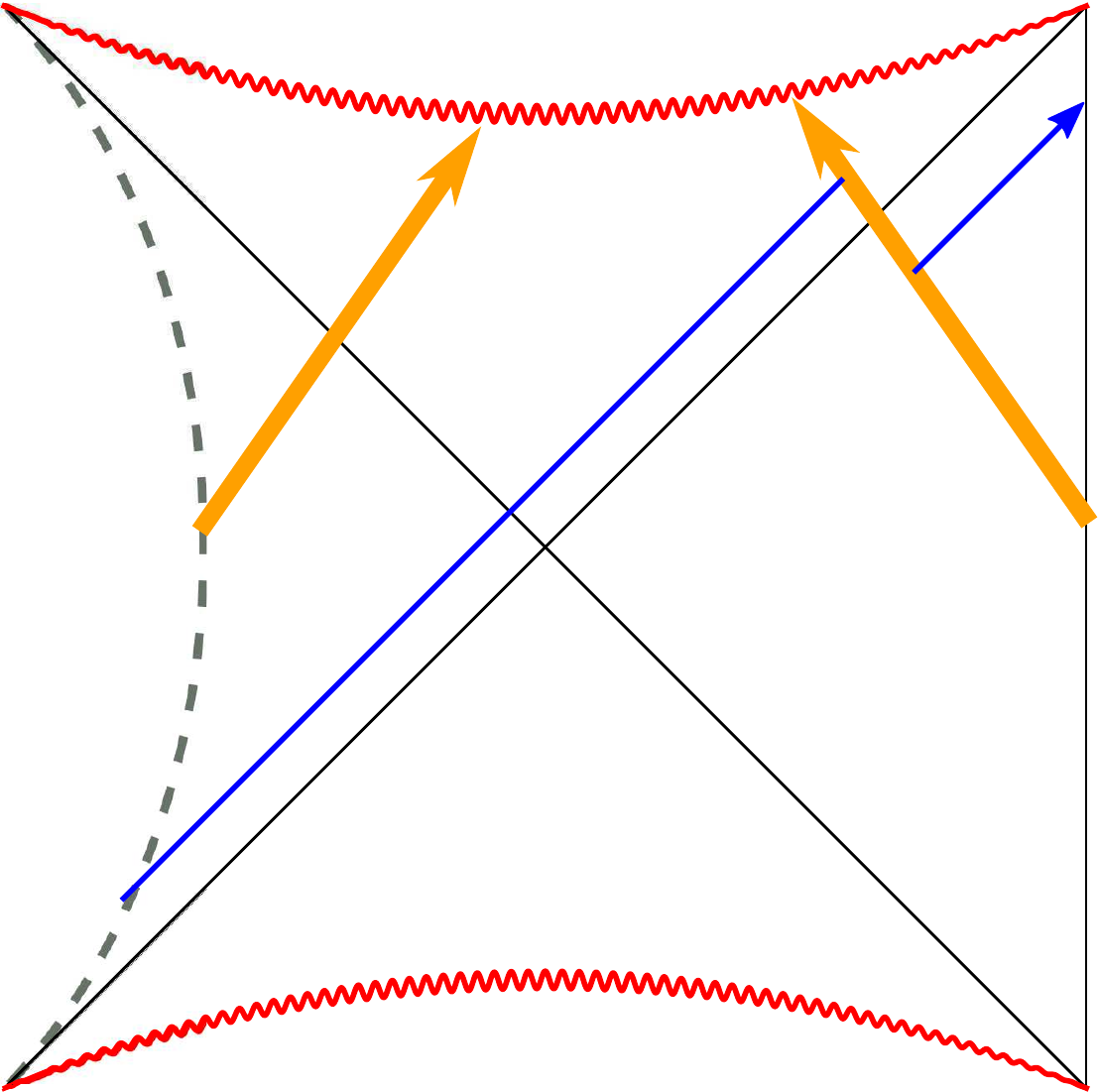}
\caption{Conjectured Penrose diagram of a typical black hole microstate, to leading order in $1/N$, with a probe created by mirror operators (blue) and two negative energy shockwaves (orange).}
\label{figureone}
\vspace{-20pt}
\end{center}
\end{figure}	
Taking into account the mirror operators, it is natural to conjecture that the geometry dual to a typical state contains not only the exterior, but also the black and white hole interiors, as well as part of the left region. This was recently emphasized in \cite{Papadodimas:2017qit}. However, we do not expect to be able to describe the full left asymptotic region due to the restriction in the frequencies $|\omega| < \omega_*$. This restriction introduces an {\it effective} cutoff of the left region, whose
nature will be described more precisely in \cite{long}.

We consider the mirror operators as gravitationally dressed with respect to the right. If we call $M,\widetilde{M}$ the mass
of the solution measured on the right, left respectively, the first law \cite{Bardeen:1973gs,Wald:1993nt,Iyer:1994ys, Hollands:2012sf,Jafferis:2015del} applied to the two-sided Cauchy slice $\Sigma$ up to the left
cutoff implies
\be
\delta M - \delta \widetilde{M} = \delta K_{\rm bulk}^{\rm full},
\ee
where $K_{\rm bulk}^{\rm full} = \int_\Sigma * (\xi T_{\rm bulk})$ and $\xi$ is the Killing vector field. This can naturally be split into the right and left contributions $K_{\rm bulk}^{\rm full}= K - \widetilde{K}$.
Since the operators are right-dressed, we have $\delta \widetilde{M} = 0$. This means
that in the code subspace the CFT Hamiltonian acts as
\be
H =M=E_0 +  K_{\rm bulk}^{\rm full},
\ee
where $E_0$ is the energy of $\rsz$.

We are now ready to set up the one-sided analogue of the double trace deformation protocol of \cite{Gao:2016bin}, which will allow us to extract particles from behind the horizon. 

A particle in the left region can be created in two ways. The first way is to actively perturb the CFT Hamiltonian at time $-t$ by a ``mirror-quench'' $\widetilde{\phi}(-t)$.
The perturbation by $\widetilde{\phi}(-t)$ creates a probe in the bulk indicated by the blue line in the figure 1 \footnote{Some care has to be taken about the interpretation of the bulk geometry to subleading order in $1/N$ \cite{long}.}. Without other perturbations the probe would end up in the singularity. The second way to create a particle in the left region is to consider a non-equilibrium state of the form
\be
U(\widetilde{\phi}) \rsz = e^{-{\beta H \over 2}} U(\phi)e^{\beta H \over 2} \rsz.
\ee
These states were extensively discussed in \cite{Papadodimas:2017qit}. For definiteness, we will consider the first scenario.

After creating an excitation in the left region by perturbing the CFT with $\widetilde{\phi}(-t)$, we perturb the CFT Hamiltonian by $e^{i g V}$, where $V= {\cal O} (0) \tO(0)$.  With the appropriate choice of the sign of $g$, this creates two negative energy shockwaves as indicated in figure 1. When analyzing the trajectory of the probe in the region around $t=0$, one should take into account the effect of the gravitational dressing of the $\tO$ operators creating the shockwave \cite{long}. Eventually,
the probe particle intersects the right negative energy shockwave and thus undergoes a time-advance. This allows it to escape the horizon and to come out in the right region, where it can finally be detected by $\phi(t)$. This is captured by the correlator
\be
\label{correlatoraaa}
C' \equiv   
 \lsz [\tphi(-t),  e^{-i g V} \phi(t) e^{i g V}]\rsz.
\ee

The conjectured bulk geometry of figure \ref{figureone} predicts that this correlator should show a sharp signal at $t \approx {\beta \over 2\pi}\log S$, similar to that of \eqref{tfdcor}. The presence of a signal of the expected form in the CFT correlator \eqref{correlatoraaa} is thus a necessary (though not sufficient) condition that the conjectured bulk geometry is the one described above and that the horizon is smooth.

\section{IV. Comparison with two-sided case}

We will now argue that the correlators $C$ and $C'$ are the same in the large $N$ limit and that, therefore, the traversability of the two-sided black hole provides evidence for the smoothness of the horizon
of the one-sided black hole. The argument is as follows. Operators in the TFD state obey the relations
\begin{align}
 \begin{split}
  {\cal O}_{\:\!\! L,\omega} &\tfd = e^{-{\beta \hat{H} \over 2}} {\cal O}_{\:\!\!R,\omega}^\dagger e^{\beta \hat{H}\over 2}\tfd  , \\ \label{tfdprop}
{\cal O}_{\:\!\!L,\omega} {\cal O}_{\:\!\!R,\omega_1}...{\cal O}_{\:\!\!R,\omega_n} &\tfd = {\cal O}_{\:\!\!R,\omega_1}...{\cal O}_{\:\!\!R,\omega_n} {\cal O}_{\:\!\!L,\omega} \tfd, \\
 [\hat{H},{\cal O}_{\:\!\!L,\omega}] {\cal O}_{\:\!\!R,\omega_1}...{\cal O}_{\:\!\!R,\omega_n} &\tfd = \omega {\cal O}_{\:\!\!L_\omega} {\cal O}_{\:\!\!R,\omega_1}...{\cal O}_{\:\!\!R,\omega_n} \tfd.
 \end{split}
\end{align}
where $\hat{H}\equiv H_R - H_L$.

Now, we consider the correlator $C$ defined in \eqref{tfdcor} and we convert all left-CFT operators into right-CFT operators by repeatedly using the equations above. As a result, the correlator $C$  takes the form
\be
C = \tfdbra {\cal X} (\phi_R,{\cal O}_R) \tfd,
\ee
where ${\cal X}$ is some time-dependent  expression involving only right-CFT operators. We can also write this as
\be
C = {1\over Z} {\rm Tr}[e^{-\beta H} {\cal X} (\phi,{\cal O})],
 \ee
  where we have dropped the subscript $R$.

We now consider the correlator $C'$ defined in \eqref{correlatoraaa} for the one-sided black hole. We follow a similar procedure by using equations \eqref{mirrordef} to convert
all mirror operators into normal operators. The important point now is that by comparing equations \eqref{mirrordef} and \eqref{tfdprop} we will get {\it exactly} the same string ${\cal X}$, i.e.
\be
C' = \lsz {\cal X} (\phi,{\cal O}) \rsz.
\ee
We have thus reduced the question about the smoothness of the horizon of a one-sided black hole, to a specific question about CFT expectation values
of ordinary (non-mirror) CFT operators. In particular, smoothness requires the proximity of the expectation value of ${\cal X}(\phi,{\cal O})$ in  the thermal ensemble ${e^{-\beta H} \over Z}$ and  a typical pure state $\rsz$. 
This is a well-defined CFT question which can in-principle be answered. It is important to notice that this condition needs to hold only for modes with $|\omega|<\omega_*$.

\section{V. A Conjecture}

We conjecture that in the large $N$ limit, and for modes with $|\omega|<\omega_*$ we have
\be
\label{conjecture}
\lim_{N\rightarrow\infty} C'= \lim_{N\rightarrow \infty}C.
\ee
As discussed above, this would provide evidence for the smoothness of the one-sided horizon.

The first step towards motivating \eqref{conjecture} is to notice that based on  general arguments, expectation values on typical pure states differ by $e^{-S}$ from those in the microcanonical ensemble $\rho_m$ of small energy spread \cite{lloyd}. Hence, we have
\be
C ' = {\rm Tr}[\rho_m {\cal X}(\phi,{\cal O})] + O(e^{-S}).
\ee
This means that to establish \eqref{conjecture}, we need to compare the expectation value of ${\cal X}$ in the canonical and microcanonical ensembles
\begin{align}
 C &= {1\over Z} {\rm Tr}[e^{-\beta H} {\cal X}(\phi,{\cal O})], \cr
C'' &\equiv {\rm Tr}[\rho_m {\cal X}(\phi,{\cal O})].
\label{ensconj}
\end{align}
These two correlators are sensitive only to the diagonal matrix elements of ${\cal X}$, since both ensembles in \eqref{ensconj} are diagonal in the energy basis, 
\be
\langle E_i|{\cal X}| E_j\rangle = f(E_i) \delta_{ij} + R_{ij}.
\ee
The Eigenstate Thermalization Hypothesis (ETH) \cite{srednicki1999approach} postulates that for ``simple observables'' the diagonal elements $f(E)$ are smooth functions of the energy and that they vary slowly with the energy, in particular ${df \over dE} \sim O(1/S)$. This suggests that the expectation values $C,C''$ differ by $1/S$ corrections --- which would imply our desired relation \eqref{conjecture}.
However, the observable ${\cal X}$ consists of products of simple operators localized at very different times of the order of scrambling time $\frac{\beta} {2 \pi} \log  S$ and it is not obvious that the ETH will hold for such observables. 

In particular, the interesting effect we want to see in the correlators $C,C''$, is coming from certain $1/S$ corrections, which get enhanced by exponential factors $e^{{2\pi \over \beta}t}$ and which become of $O(1)$ at scrambling time. Hence, the non-trivial content of the conjecture \eqref{conjecture} is that these ``chaos-enhanced'' $1/S$ corrections are the same in the two ensembles. Relatedly, it suggests that if some operators obey ETH, their product will also obey ETH, even when the operators are widely separated. 

Some evidence for this conjecture follows from the observation that  the ETH is expected to be robust under multiplication of operators \cite{Marolf:2013dba,long}, at least for small separations in time between operators and for extremely large time separations, where the matrix elements become almost totally uncorrelated. It is natural that the same is true for intermediate times, which include times of the order of scrambling time.
 
\begin{figure}[!t]
\begin{center}
\begin{subfigure}[l]{.23\textwidth}
\includegraphics[width=\textwidth]{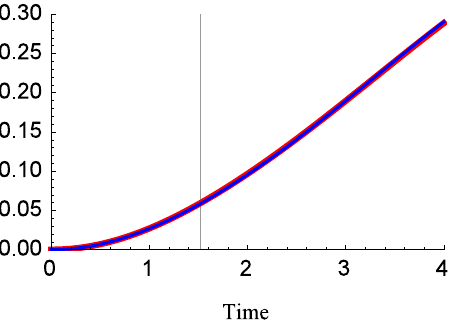}
\end{subfigure}
\begin{subfigure}[r]{.23\textwidth}
\includegraphics[width=\textwidth]{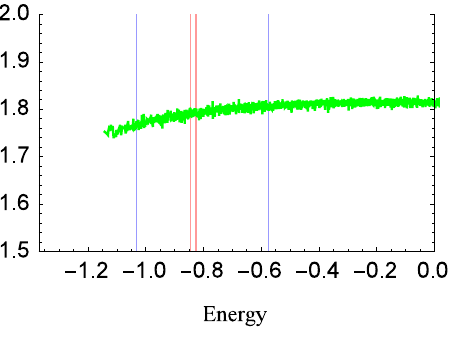}
\end{subfigure}
\end{center}
\caption{\small Numerics in SYK model for $N=24$. Left : $ \langle  \{\psi^I(t),\psi^J(0)\}^2\rangle$ in  thermal (blue) and typical pure state (red). The scrambling time is designated by the vertical line. 
Right: diagonal matrix elements $\langle E_i| \{\psi^I(t),\psi^J(0)\}^2|E_i\rangle$ for $t\approx$ scrambling time. It shows slow variation with the energy, compatible with the ETH, within the dominant regions of the canonical ensemble (blue) and the microcanonical ensemble (red). }
\label{figuretwo}
\end{figure}

Further evidence can be found by considering some simple models. Firstly, in large $c$ 2d CFTs with sparse spectrum, it was
argued in \cite{Turiaci:2016cqc} that the commutator of two operators separated by times of the order of the scrambling time is dominated by certain time ordered terms. This can be used to transform the out-of-time-order correlator to a time-ordered correlator, for which it is generally assumed that factorization is still applicable and that would imply our conjecture. Secondly, if we assume that the correlators are dominated by the Virasoro identity block even at scrambling time, then there is evidence \footnote{J. Sonner, private communication}
 that  \eqref{conjecture} follows. 

Finally, we have performed some numerical studies in the SYK model, of computing out-of-time-order correlators at scrambling time \footnote{Unfortunately for the values of $N$ that we were able to study numerically, there is no large separation between thermalization and scrambling times.}, both in the thermal ensemble and in typical pure states, where we find  good agreement as shown in figure \ref{figuretwo}.

\section{VI. An analogue of Hayden-Preskill}

We observe that the mirror operators $\tO$ discussed in the previous sections realize an analogue of the Hayden-Preskill protocol, in the form described in \cite{Maldacena:2017axo}. We start with a black hole in AdS dual to a microstate $\rsz$. At some time
$t_0 \approx - t_S$  (here $t_S$ is scrambling time), we throw a qubit from the boundary into the black hole. This qubit is created in the bulk by acting with the CFT operator $U_\epsilon = e^{i \epsilon \phi(t_0)}$ (appropriately smeared). We wait until the particle has been absorbed, and then we ask what is the CFT operator we need  to measure in order to extract the quantum information of the qubit. 

One natural way to do this, is by perturbing the CFT Hamiltonian by an interaction of the form $V= {\cal O}(0) \tO(0)$ with an appropriate coupling constant, which produces two negative energy shockwaves. The infalling particle
collides with one of the shockwaves (the ``mirror shock'') and undergoes a time-advance, pushing it into the left region. 
It can then be measured by the mirror operator $\widetilde{\phi}(t_S)$. The result of this measurement is captured by a correlator similar to \eqref{correlatoraaa}, with the roles of $\phi$ and $\tilde{\phi}$ reversed.
The conjecture of the previous section implies that this correlator can  extract the quantum information of the probe.

\begin{figure}[!b]
\begin{center}
\includegraphics[width=.4\textwidth]{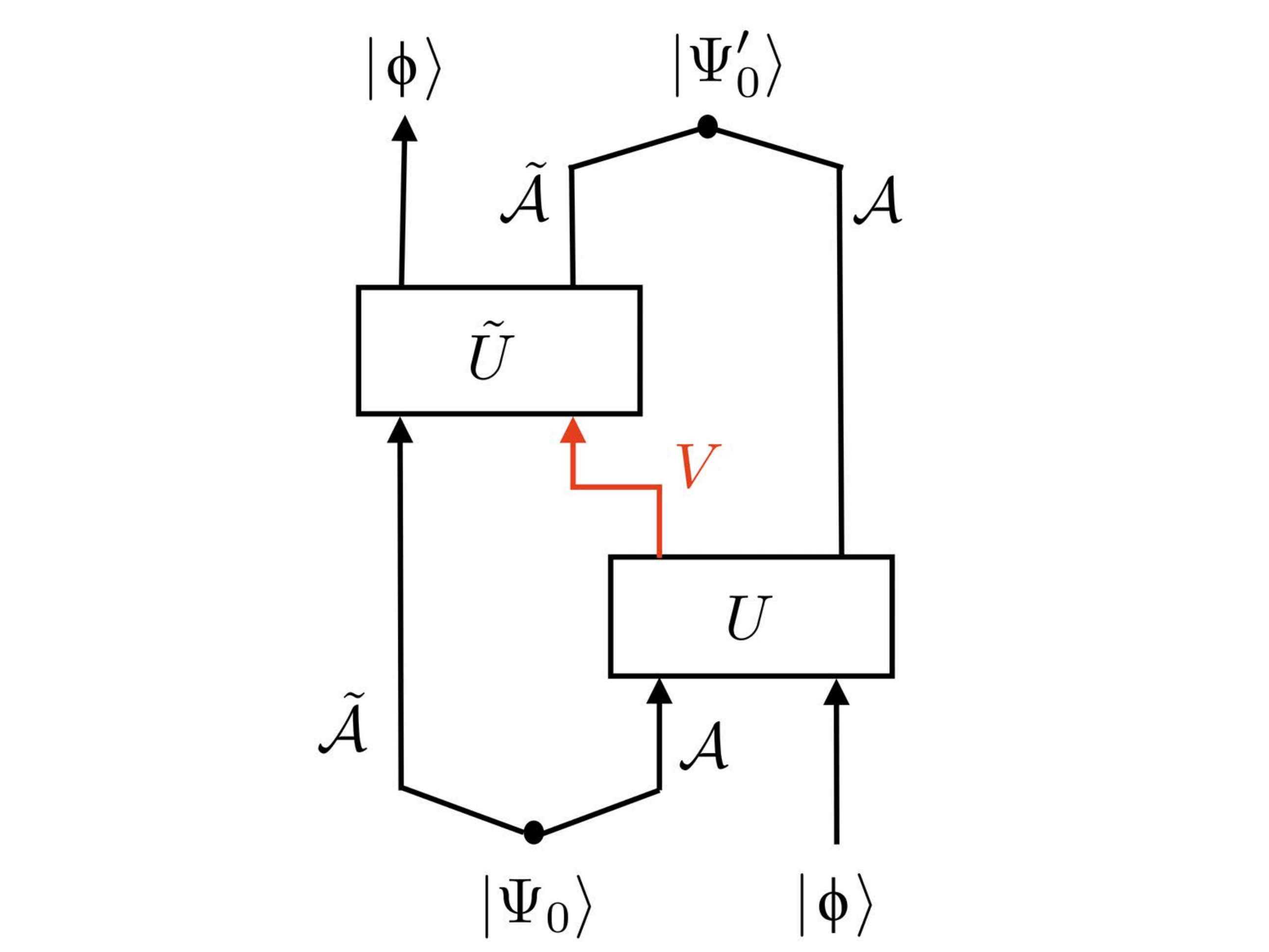}
\end{center}
\caption{\small A realization of the Hayden-Preskill protocol: the code subspace approximately factorizes into a tensor product corresponding to the algebras ${\cal A}$,${\cal A'}$. These tensor factors are entangled and provide the reservoir of EPR pairs needed to perform the teleportation. Here $U,\widetilde{U}$ is time evolution in the CFT and $V = {\cal O} \tO$ denotes the perturbation of the CFT Hamiltonian.}
\end{figure}

The Hayden-Preskill protocol can only be applied after the half-point of evaporation, when the black hole is maximally entangled with the early radiation. The analogue statement in our case is that in order to define the operators $\tphi,\tO$, one needs to have knowledge of the microstate, as the $\tphi,\tO$'s are state-dependent operators. We remind the reader that even in the original Hayden-Preskill protocol, the decoding operation is state-dependent.

\section{VII. Comments}

We formulated a necessary condition for the smoothness of the horizon of a typical black hole microstate in terms of CFT correlators of local operator at scrambling time.
 We argued that smoothness of the horizon requires that these correlators are similar in the canonical and microcanonical ensembles. We provided some preliminary evidence in favor of this conjecture. 
These observations imply that for certain purposes it is meaningful to consider part of the left region of the extended AdS-Schwarzchild geometry.

Our arguments made use of state-dependent operators.
Since the boundary observer has in principle unlimited resources, this fits within the conventional framework of quantum mechanics. Indirectly, this provides evidence for the relevance of the state-dependent operators for the infalling observer.
Further details and open questions will be discussed in an upcoming longer article \cite{long}.

\vspace{0.3cm}
\section{Acknowledgements}

We would like to thank G. Gur-Ari, D. Harlow, D. Jafferis, E. Kiritsis, J. Maldacena, L. Motl, R. Myers, S. Raju, J. Sonner, H. Verlinde, S. Wadia for useful discussions.  RvB would like to thank NCCR SwissMAP of the Swiss National Science Foundation for financial support. SFL thanks The Netherlands Organization for Scientific Research (NWO) for financial support and CERN for hospitality during the completion of this work.  KP would like to thank KNAW and NWO for support and the University of Crete for hospitality.

\bibliographystyle{apsrev4-1} 
\bibliography{references.bib} 

\end{document}